\documentstyle[pra,aps]{revtex}
\def\up{\uparrow}
\def\dn{\downarrow}

\begin{document}
\preprint{\today}
\draft
%
%
\title{Measured Quantum Dynamics of a Trapped Ion}
\author{Lorenza Viola and Roberto Onofrio}
\address{Dipartimento di Fisica ``G. Galilei,'' 
Universit\`a di Padova, Via Marzolo 8, Padova, Italy 35131}
\date{\today}

\maketitle
%
%
\begin{abstract}
The measurement process is taken into account in the dynamics of trapped 
ions prepared in nonclassical motional states. The induced decoherence is 
shown to manifest itself both in the inhibition of the internal 
population dynamics and in a damping of the vibrational motion 
without classical counterpart. 
Quantitative comparison with present experimental 
capabilities is discussed, leading to a proposal for the verification of 
the predicted effects.
\end{abstract}
%
%
%
\begin{center}
(Preprint {\sc dfpd/gp/53}, to be published in {\sl Physical Review A, Rapid Communications })
\end{center}
%
Quantum decoherence is commonly understood as the decay of the off-diagonal
density matrix elements responsible for the nonclassical properties of 
superposition states. 
Its key role in the transition from quantum to classical 
physics has been lately recognized, 
underlying the reason we are prevented from everyday observation of 
quantum features in the macroscopic realm  \cite{ZUREK1,ZUREK2}. 
Decoherence is usually conspiring against stable preparation of quantum 
states through different mechanisms. Despite of the fact that some of them, 
for instance of thermal or instrumental nature, can be quenched by proper 
technological developments, an {\sl ultimate} source remains 
due to the coupling between the observed system and the meter. 
This coupling has in principle to be taken into account whenever 
physical information is extracted from the system.

Several proposals have been discussed so far for experimentally testing the
issue of decoherence, among these the ones employing superposition states in 
molecules or crystals \cite{JANS}, superconducting rings \cite{LEGGETT}, 
nonlinear Kerr media \cite{YURKE} and cavity QED setups 
\cite{PHOENIX,HAROCHE}. Recently, a further scenario has been disclosed 
succeeding in the generation and detection of nonclassical motional states 
of single trapped and cooled ions \cite{MEEKHOF}, 
also opening the way to the creation of mesoscopic Schr\"odinger cat 
states and the controlled study of their subsequent death \cite{MONROE96}.
The possibility to manipulate the internal and external state of the ion 
by means of suitably arranged laser beams, after cooling down to zero-point 
vibrational energy \cite{MONROE95}, and the extremely low dissipation
attainable on the vibrational motion itself, have been crucial in 
determining the success of this technique.

It is the purpose of this paper to investigate the decoherence introduced 
by a continuous measurement process on the dynamics of a trapped and 
laser-irradiated ion.
It turns out that peculiar quantum mechanical effects are in principle 
detectable in the evolution of both electronic and motional degrees of freedom 
of the trapped ion.

The starting point of our analysis is the Lindblad equation for 
the density operator $\hat{\rho}(t)$ of a system subjected to a continuous 
measurement of the generic observable $\hat{A}$ 
\cite{LINDBLAD,CARMICHAEL,MQM},
\begin{equation}
{d \hat{\rho}(t) \over dt}= - {i \over \hbar} 
[\hat{H}_0+\hat{H}_{int}(t),\hat{\rho}(t)] - {\kappa \over 2} 
[\hat{A},[\hat{A},\hat{\rho}(t)]]\:.   
\label{MASTER}
\end{equation}
The first contribution in the right-hand side of (\ref{MASTER}) describes the 
dynamics of the unmeasured system. The Hamiltonian operator $\hat{H}_0$ 
accounts for the free motion of the vibrational and electronic degrees of 
freedom of the ion,
\begin{equation}
\hat{H}_0= \hat{H}^{cm}+\hat{H}^{el}=\hbar \omega 
\, \hat{a}^{\dagger} \hat{a} + \hbar \omega_{21} \,\hat{\sigma}_z \:,
\label{HZERO}
\end{equation}
being $\omega$ the frequency of the harmonic trap, hereafter assumed to 
be much larger than the atomic radiative linewidth (strong confinement 
limit, $\omega \gg \Gamma$), 
$\omega_{21}$ the transition frequency of the two-level system, 
$\hat{a}$ and $\hat{\sigma}_z$ respectively the annihilation operator of 
the 1D boson mode and the atomic pseudo-spin $\hat{z}$-operator. 
Coupling between internal and translational dynamics is achieved through a 
nonlinear multiquantum Jaynes-Cummings Model (JCM) interaction, which in 
rotating-wave approximation is written as \cite{VOGEL94,VOGEL95}
\begin{equation}
\hat{H}_{int}(t)={\hbar \Omega_0 \over 2} \,\mbox{e}^{i \omega_L t} \cos 
[\eta (\hat{a}+\hat{a}^{\dagger}) + \varphi] \, \hat{\sigma}_- + 
\mbox{H.c.}  \:, \label{HINT}
\end{equation}
where $\omega_L$ is the laser operating frequency, $\Omega_0$ the 
fundamental Rabi frequency, $\eta=(\omega_L/c)(\hbar/2 m \omega)^{1/2}$ 
the Lamb-Dicke parameter related to the standard quantum limit for the 
localization of the ion and $\hat{\sigma}_-$ the electronic lowering 
operator. In Eq. (\ref{HINT}) a standing-wave laser field is considered, 
the phase $\varphi$ determining the position of the trap potential with 
respect to the wave. If, in particular, the ion is trapped at the node 
of the light field ($\varphi = \pm \, \pi/2$) and $\eta \ll 1$ 
(Lamb-Dicke limit), the well-known linear one-quantum JCM or anti-JCM 
operators $\eta(\hat{a}\hat{\sigma}_+ + \hat{a}^{\dagger}\hat{\sigma}_-)$,   
$\eta(\hat{a}^{\dagger} \hat{\sigma}_+ + \hat{a}\hat{\sigma}_-)$   
may be recovered \cite{JAYNES,RISKEN,CIRAC}. 
The second contribution in Eq. (\ref{MASTER}) is responsible for the opening 
of the dynamics due to the effect of the measurement. The parameter 
$\kappa$, supposed to be time-independent, 
expresses the strength of the coupling to the measuring apparatus \cite{MQM}. 
The choice of the measured observable $\hat{A}$ clearly depends upon the 
specific experimental procedure. 
If the occupancy of the internal ground state $\dn$ is 
monitored, for instance by collecting the fluorescence light emitted after 
stimulated transitions to a third auxiliary level as done in \cite{MEEKHOF},
the appropriate operator in (\ref{MASTER}) is given in the pseudo-spin 
formalism by $\hat{A}=\hat{\sigma}_-\hat{\sigma}_+$. Any other admissible 
observable of the system could be analyzed in principle. 

It is convenient to project Eq. (\ref{MASTER}) in the representation 
defined by the eigenkets of $\hat{H}_0$,  
$|S,n \rangle$, $S=\dn,\up$, $n=0,\ldots, \infty$. The analysis is greatly
simplified if, as in practical conditions \cite{MEEKHOF,MONROE95}, 
the ion is in the low excitation regime ($\omega \gg \Omega_0$) and the laser 
is tuned to the $k$-{th} vibrational sideband, 
$\omega_L=\omega_{21} \pm k \omega$, with $+$ $(-)$ corresponding to blue 
(red) respectively. Thus transitions involving the exchange of $k$ vibrational 
quanta are resonantly enhanced, while the off-resonant couplings rapidly 
oscillating at frequency $\omega$ can be disregarded, the so-called JCM 
approximation \cite{VOGEL94,VOGEL95}. By choosing, without loss of generality,
a $k$-th blue resonance, transitions between states $|\dn,n \rangle$, 
$|\up,n+k \rangle$ are driven and the master equation (\ref{MASTER}) leads to 
the following closed set of equations
\begin{eqnarray}
 & & \hspace*{-1mm} \left\{ \begin{array}{ll}
 \dot{\rho}_{\dn n \dn m}(t) & \hspace{-1mm}
= -i \omega(n-m)\rho_{\dn n\dn m}(t)
    - i/2 \left[ \Omega_{n n+k} \rho_{\up n+k\dn m}(t) \mbox{e}^{i \omega_L t}
    - \Omega_{m m+k} \rho_{\dn n\up m+k}(t) \mbox{e}^{-i \omega_L t}
 \right] \:, \\
 \dot{\rho}_{\up n+k \up m+k}(t) & \hspace{-1mm}
= -i  \omega(n-m)\rho_{\up n+k\up m+k}(t)
  - i/2 \left[ \Omega_{n n+k} \rho_{\dn n\up m+k}(t) \mbox{e}^{-i \omega_L t}
     - \Omega_{m m+k} \rho_{\up n+k\dn m}(t) \mbox{e}^{i \omega_L t}
 \right] \:, \\
 \dot{\rho}_{\dn n \up m+k}(t) & \hspace{-1mm} = 
\left[ -i\left( \omega(n-m)-\omega_{21} \right) -{\kappa/ 2} \right] 
   \rho_{\dn n\up m+k}(t) - i/2 \left[ \Omega_{n n+k} \rho_{\up n+k\up m+k}(t) 
   - \Omega_{m m+k} \rho_{\dn n\dn m}(t) \right]
   \mbox{e}^{i \omega_L t} \:, \\
 \dot{\rho}_{\up n+k \dn m}(t) & \hspace{-1mm} = 
\left[ -i\left( \omega(n-m)+\omega_{21} \right) -{\kappa / 2} \right] 
     \rho_{\up n+k\dn m}(t) - i/2 \left[ \Omega_{n n+k} \rho_{\dn n\dn m}(t) 
     - \Omega_{m m+k} \rho_{\up n+k\up m+k}(t) \right]
     \mbox{e}^{-i \omega_L t} \:, 
\end{array} \right.
\label{EQS}
\end{eqnarray}
where, as customary, the nonlinear $k$-quantum Rabi frequencies have been
introduced \cite{VOGEL95}, 
\begin{equation}
 \Omega_{n n+k}=\Omega_0 \langle n |\, \cos[ \eta 
(\hat{a}+\hat{a}^{\dagger})+ \varphi] \, |n+k \rangle \:.
\label{RABI}
\end{equation} 
Eqs. (\ref{EQS}) can be analitically solved through standard techniques, 
the details of which will be reported in a future extended paper. Two special 
cases can be easily handled. When $\kappa=0$ the solution of Eqs. (\ref{EQS}) 
is straightforwardly obtained by solving the Schr\"odinger equation for the 
closed system; for $\kappa >0$ and $n=m$ a two-level measured evolution is 
instead recognizable within each Jaynes-Cummings manifold 
$ \{ |\dn,n \rangle$,$|\up, n+k \rangle \} $. 
In analogy to the situation already investigated in \cite{MQM}, a Rabi-like 
oscillatory behavior with frequency $\Omega_{n n+k}$ or a Zeno-like 
dynamically frozen regime are then expected for values of $\kappa$ 
respectively smaller or larger than the critical one 
$\kappa^{crit}_{n n+k}=4 \Omega_{n n+k}$. Starting from the general solution 
of Eqs. (\ref{EQS}), the properties of the internal and vibrational motions 
can be studied by considering the reduced density matrices
\begin{equation}
\sigma_{SS'}(t)= \sum_{n=0}^{\infty} \rho_{Sn,S'n}(t)\:, \hspace{1cm}
\rho^{cm}_{nm}(t)= \sum_{S=\dn,\up} \rho_{Sn,Sm}(t)\:.
\label{REDUCED}
\end{equation}
As in the experiment reported by Meekhof {\sl et al.} \cite{MEEKHOF}, we 
will assume that at initial time no entanglement between internal and 
translational degrees of freedom is present and only the lower electronic 
level is populated, {\sl i.e.} 
$ \rho_{Sn,S'm}(0)=\sigma_{SS'}(0)\, \rho^{cm}_{nm}(0)=
  \delta_{SS'} \delta_{S \dn}  \, \rho^{cm}_{nm}(0)$. 

Let us first focus on the effect of the measurement on the internal 
subdynamics. The measured occupancy of level $\dn$, 
$P_{\dn}(t)=\mbox{Tr}(\hat{\sigma}(t) \hat{\sigma}_-\hat{\sigma}_+)=
\sigma_{\dn\dn}(t)$ is found to be
\begin{equation}
P_{\dn}(t)={1 \over 2} \bigg\{ 1+ \mbox{e}^{-(1/4) \kappa t}
\sum_{n=0}^{\infty} \rho^{cm}_{nn}(0) \bigg[ 
\cos(w_{nn}t)+{\kappa \over 4 w_{nn}} \sin(w_{nn}t) 
\bigg] \bigg\}  \:,      \label{PDOWN}
\end{equation}
where the diagonal center-of-mass matrix elements characterize the 
number state distribution of the vibrational motion and the following 
conventions are introduced, for future reference, for the frequencies
\begin{equation}
w_{nm}  =  \sqrt{ \bigg( {\frac{\Omega_{n n+k} + \Omega_{m m+k}}{2}\bigg)}^2
-{\kappa^2 \over 16} }  \:, \hspace{1cm}
u_{nm}  =  \sqrt{ \bigg( {\frac{\Omega_{n n+k} - \Omega_{m m+k}}{2}\bigg)}^2
-{\kappa^2 \over 16} }  \:.  
\label{SYMB}
\end{equation}
Eq. (\ref{PDOWN}) has to be compared with the analysis presented in Ref. 
\cite{MEEKHOF}. Apart from a factor 2 entering the definition of the 
multiquantum frequencies (\ref{RABI}), it is worth to emphasize that the decay 
rate of the signal (\ref{PDOWN}), determined by the value of $\kappa$, is 
independent upon the vibrational quantum number, corroborating a different 
origin of the $n$-increasing decoherence instead observed in \cite{MEEKHOF} 
and there ascribed to laser intensity noise and trap drive instabilities.
In the simplest case of an initial Fock state, 
$\rho^{cm}_{nn}(0)=\delta_{n \overline{n}}$, Eq. (\ref{PDOWN}) simplifies to
\begin{equation}
P_{\dn}^{Fock}(t)={1 \over 2} \bigg\{ 1+ \mbox{e}^{-(1/4) \kappa t}
\bigg[ \cos(w_{\overline{n}\,\overline{n}}t)+
{\kappa \over 4 w_{\overline{n}\,\overline{n}} } 
\sin(w_{\overline{n}\,\overline{n}}t ) 
\bigg] \bigg\}  \:.      \label{FOCK}
\end{equation}
In the Zeno-like regime, $\kappa > \kappa^{crit}_{\overline{n}\,
\overline{n}}$, the frequency $w_{\overline{n}\,\overline{n}}$ becomes 
imaginary and an almost complete inhibition of the evolution occurs after a 
transient of the order $\kappa^{-1}$. Moreover, according to (\ref{FOCK}), 
even in the opposite weak coupling limit 
$0< \kappa \ll \kappa^{crit}_{\overline{n}\,\overline{n}+k}$ 
the measurement induced decoherence appears  through three distinct 
signatures: an exponential decay of the signal amplitude, a frequency 
shift of the pseudo-oscillation and the presence of a sinusoidal term. 
More elaborated examples are available by changing the initial vibronic 
distribution in (\ref{PDOWN}), as discussed in \cite{ONVIO96} with emphasis on 
the collapses and revivals arising from a coherent initial state.

Until now an effect of the measurement has been identified on the dynamics of 
the internal degree of freedom, a fact which is quite expected since the 
latter is {\sl directly} related to the observed quantity $\hat{A}$. However, 
the back-action of the measurement on the evolution of the coupled vibrational 
motion has likewise to be examined. The motional degree of freedom also feels 
the measuring apparatus, originating {\sl indirect} decoherentization of the 
reduced density matrix $\rho^{cm}_{nm}(t)$ introduced in (\ref{REDUCED}) 
and therefore modifying the average values of any  center-of-mass observable, 
$\langle \hat{\cal O}^{cm}(t) \rangle = 
\mbox{ Tr}(\hat{\rho}^{cm}(t)\hat{\cal O}^{cm} )$. We restrict ourselves here 
to two representative cases, center-of-mass position and energy. They can be 
shown to evolve respectively as
\begin{eqnarray}
\langle \hat{x}^{cm}(t) \rangle & = & 
\sqrt{{\hbar \over 2 m \omega}}  \mbox{e}^{-(1/4)\kappa t} \,
\sum_{n=0}^{\infty}\,  \Re\mbox{e}\bigg\{ \mbox{e}^{i \omega t}
\rho^{cm}_{n n+1}(0)  
\bigg[
(\sqrt{n+k+1} + \sqrt{n+1}) 
\Big[ \cos(u_{n n+1}t) + {\kappa \over 4 u_{n n+1}} \sin(u_{n n+1}t) 
\Big] \nonumber \\ 
& - & (\sqrt{n+k+1} - \sqrt{n+1}) 
\Big[ \cos(w_{n n+1}t) + {\kappa \over 4 w_{n n+1}} \sin(w_{n n+1}t) 
\Big]    \bigg] \bigg\} \:,
\label{POSITION}
\end{eqnarray} 
\begin{equation}
\langle \hat{H}^{cm}(t) \rangle = \hbar \omega 
\bigg( \overline{n} + {1 \over 2} + {k \over 2} \bigg) -
{\hbar \omega \over 2}\,k\, \mbox{e}^{-(1/4)\kappa t}
\sum_{n=0}^{\infty} \, \rho^{cm}_{nn}(0)  
\Big[ \cos(w_{n n}t) + {\kappa \over 4 w_{n n}} \sin(w_{n n}t) 
\Big] \:,      \label{ENERGY}
\end{equation}
being $u_{n n+1}$, $w_{n n+1}$ and $w_{n n}$ previously defined in (\ref{SYMB}).
Eqs. (\ref{POSITION}) and (\ref{ENERGY}) contain the main result of this 
paper, implying that the average position and energy exponentially 
relaxe to a constant value with a rate $\kappa/4$, a {\sl quantum damping} 
purely due to the measurement process, as already introduced 
for the case of a single degree of freedom \cite{ONVIO}. 
More generally, quantum damping can be proven to occur for any observable
$\hat{\cal O}^{cm}$, leading to the asymptotic behavior 
\begin{equation}  
\lim_{t \rightarrow \infty} \, \langle \hat{\cal O}^{cm}(t) \rangle = 
\frac{1}{2} 
\sum_{n=0}^{\infty} \,
\Big( \langle n|\hat{\cal O}^{cm}|n \rangle + 
      \langle n+k|\hat{\cal O}^{cm}|n+k \rangle \Big) \, \rho^{cm}_{nn}(0) \:. 
\label{DAMPING}
\end{equation}
As in the previously discussed decoherence effect on $P_{\dn}(t)$, only 
diagonal contributions of $\hat{\rho}^{cm}(0)$ enter Eq. (\ref{ENERGY}), 
whereas the existence of the damping for the position $\hat{x}^{cm}$ (as well 
for any other negative-parity observable) crucially rests on the coherence of 
the initial state number occupation. Relation (\ref{DAMPING}) allows one to 
evaluate the asymptotic position variance: by choosing 
$\hat{\cal O}^{cm}=(\hat{x}^{cm}- \langle \hat{x}^{cm} \rangle )^2$, one gets 
\begin{equation}
\lim_{t \rightarrow \infty} \, \Delta \hat{x}_{cm}^2(t)  =
4\bigg( {\hbar \over 2 m \omega} \bigg) 
\bigg( \overline{n} + {1 \over 2} + {k \over 2} \bigg) \:, 
\label{VARIANCE}
\end{equation}
with $\overline{n}=\sum_{n=0}^{\infty} n \, \rho^{cm}_{nn}(0)$ the initial 
average vibrational quantum number. Hence, the position damping
(\ref{POSITION}) does not imply a parallel suppression of the position 
variance. Analogous considerations can be repeated for the center-of-mass 
momentum, enforcing the validity of the Heisenberg principle.
Moreover, center-of-mass position, momentum and energy are linked together 
via a simple asymptotic relationship,
\begin{equation}
\lim_{t \rightarrow \infty} \, \bigg\langle 
{\hat{H}^{cm}(t) \over 2} \bigg\rangle = 
\lim_{t \rightarrow \infty} \, \bigg\langle 
{m \omega^2 \hat{x}^2_{cm}(t) \over 2} \bigg\rangle = 
\lim_{t \rightarrow \infty} \, \bigg\langle 
{\hat{p}^2_{cm}(t) \over 2 m} \bigg\rangle  \:,
\end{equation}  
pointing out that energy equipartition is preserved when quantum damping is 
in action. Eq. (\ref{ENERGY}) deserves some further comments. First, the 
asymptotic average energy  corresponds to heating of the vibrational motion 
with respect to the initial value 
$\langle \hat{H}^{cm}(0) \rangle = \hbar \omega (\overline{n} + 1/2)$, which is 
reasonable due to the blue sideband driving. In the case $k=0$ however, the 
energy is conserved for {\sl any} value of the measurement coupling $\kappa$, 
as physically understandable since the laser is tuned to the electronic 
transition and the interaction does not change the number of vibrational 
quanta. Therefore, no back-action of the measuring apparatus on the motional 
energy is present in such conditions, consistently with the recent analysis 
by de Matos Filho and Vogel where quantum nondemolition measurements of the 
vibronic energy based on a zero-quantum JCM interaction have been proposed
\cite{VOGEL96}.  

Let us conclude by briefly discussing the observability of the effects 
predicted on both the internal and external motions of the ion. As a 
preliminary remark, the ensemble formalism throughout adopted here finds its 
justification since averages of measurements on single ions each time 
identically prepared are to be handled in laboratory. Concerning the 
deformation of the two-level dynamics (\ref{PDOWN}), one has essentially to 
design a more elaborated version of the experiment performed by Itano {\sl et 
al.} to observe the quantum Zeno effect \cite{ITANO}. Compared to this, a 
complication arises due to the dependence of the critical measurement coupling 
$\kappa^{crit}_{n n+k}$ on the Jaynes-Cummings manifold, owing to which a 
sharp transition from oscillatory to overdamped regime is only expected in the 
simplest case of a Fock state (\ref{FOCK}).  
An experimental scheme aimed at testing quantum damping is slightly more 
complicated since a further measurement is required to check the predicted 
average values. This is not conceptually different from the verification of 
any quantum mechanical prediction. If, for instance, positional damping is 
considered, the procedure involves many replicas of the evolution of the whole 
system + meter up to time $t$ each followed by an instantaneous measurement of 
the ion position. By subsequently repeating the experiment for different times 
$t$, the average value $\langle x^{cm}(t)\rangle$ can be accumulated and 
compared to (\ref{POSITION}). In practice, the readout of the ion position 
could be obtained by measuring the charge induced on two auxiliary end caps in 
the trap, provided the signal-to-noise ratio is large enough \cite{BROWN}.
Alternatively, less direct methods could be adopted, such as the already 
demonstrated selective measurement of position in which fluorescence light 
induced by a localized probe beam is detected \cite{AMINOFF}. Leaving aside to 
a future analysis the detailed discussion of the detection scheme, a more 
quantitative insight of the relevant timescales can be already gained by 
assuming for the parameter $\kappa$ the same value adopted in \cite{MQM} to 
fit the data of the quantum Zeno experiment \cite{ITANO}, 
$\kappa \approx 4.9\cdot 10^4\,\mbox{s}^{-1}$. Also notice that the same 
auxiliary transition
$\mbox{}^2S_{1/2}\rightarrow \mbox{}^2P_{3/2}$ of $\mbox{}^9\mbox{Be}^+$ 
has been chosen in \cite{MEEKHOF}. The resulting decay time is 
$\tau=4 \kappa^{-1}=816\,\mu$s for both the internal and motional dynamics, to 
be compared with the decoherence reported in \cite{MEEKHOF} for a Fock state, 
namely $\tau^{exp}=84\,\mu$s for the 
$|\dn,0\rangle \rightarrow |\up, 1\rangle$ transition. Being additional 
relaxation mechanisms due to spontaneous decay of the electronic transition 
and mechanical dissipation of the ion motion negligible \cite{ITANO,BROWN}, 
these numerical estimates show that a hundredfold improvement in the stability 
performances of the experimental setup should make the overall decoherence 
dominated by the measurement induced one. 
Since a ratio $\kappa/\kappa^{crit}_{01}=2.1\cdot 10^{-2}$ is obtained in the 
example considered above, both a Rabi-like behavior in the two-level dynamics 
and an underdamped oscillation in the vibronic motion are expected. An 
increase of this ratio can be attained by lowering the fundamental Rabi 
frequency $\Omega_0$ by two orders of magnitude, allowing one to enter the 
region where the two contributions in the right-hand side of Eq. (\ref{MASTER}) 
strongly compete.
    
In summary, we have proposed a quantitative model to examine the influence of 
a continuous measurement process on the dynamics of a trapped ion. The 
decoherence induced by the interaction with the measuring apparatus has been 
identified through peculiar modifications of both electronic and vibrational 
observable properties of the ion. Detecting such effects gives further 
stimulus to deepen our knowledge of the mesoscopic world and may complement 
the direct reconstruction of the density matrix via quantum tomography of 
single trapped ions recently achieved in laboratory \cite{LEIBFRIED}.



This work has been supported by INFN, Sezione di Padova, and INFM, 
Sezione di Roma 1, Italy. 



\end{document}